\documentclass[11pt, a4paper]{article}
\usepackage{moriond,epsfig}
\usepackage{amssymb}
\usepackage{mathptmx}
\usepackage{amsbsy}   
\usepackage{amsmath}  

\def\be{\begin{equation}}
\def\ee{\end{equation}}
\def\bea{\begin{eqnarray}}
\def\eea{\end{eqnarray}}

\begin{document}
\title{HEAVY ION PHYSICS WITH THE ALICE EXPERIMENT AT LHC}

\author{Chiara ZAMPOLLI (1) for the ALICE Collaboration}
\address{(1) Museo Storico della Fisica e Centro Studi e Ricerche Enrico Fermi,
Rome, Italy, and Sezione INFN, Bologna, Italy, and Dipartimento di Fisica
dell'Universit\`a di Bologna, Italy}

\maketitle\abstracts{ALICE is the experiment at the LHC collider at CERN 
dedicated to heavy ion physics. In this report, the ALICE detector will be 
presented, together with its expected performance as far as some selected
physics topics are concerned.} %
\noindent

\section{Introduction}
\label{sec:intro}
Besides its proton-proton (pp) physics program, the Large Hadron Collider 
at CERN, 
which is going to start operating in early 2008, will deliver lead-lead (PbPb) 
collisions at a centre-of-mass energy $\sqrt{s_{NN}}$ = 5.5 TeV, 
opening the door to unexplored regimes 
in heavy-ion particle physics from several points of view, such as the 
centre-of-mass energy, the energy density reached, the lifetime 
and the volume of the Quark-Gluon Plasma (QGP) system which may be created in 
the collision. ALICE (A Large Ion Collider Experiment) will be the
experiment dedicated to study heavy-ion collisions at the LHC. Its physics program will span over a large 
number of observables, from the global properties of the collisions
(multiplicities, rapidity distributions...), to more selective
QGP signals (like direct photons, charmonium and bottomonium...). 
In this report,
an overview of the LHC and the ALICE detector will be first provided. 
Then, the physics performance of the experiment as far as open charm and 
open beauty detection, quarkonia and jet physics studies will be discussed. 
Finally, the ALICE ``First Day'' Physics will be briefly presented.  

\section{The ALICE Experiment at the LHC}
\label{sec:ALICE}
The Large Hadron Collider experimental program will deal with  
different collision systems. Apart from the main pp and PbPb runs, also 
asymmetric collisions (such as p-Pb) and collisions with heavy ions other 
than Pb (e.g. Ar) will be studied. 
Table~\ref{runningCond} quotes the expected values for the most significant 
running parameters for the pp and PbPb collision systems, which 
have been used as a reference for the results presented hereafter.

ALICE will be the experiment at the LHC dedicated
to heavy-ion physics. In the central rapidity region 
$-$0.9 $<$ $\eta$ $<$ 0.9, inside the L3 magnet 
(providing the experiment with a weak solenoidal magnetic field B = 
0.2 - 0.5 T), ALICE will be endowed with subdetectors specialized in 
tracking and identifying the particles produced in the collisions, namely
the ITS (Inner Tracking System), the TPC (Time Projection Chamber),
the Transition Radiation Detector (TRD), and the Time Of 
Flight (TOF).
The central region will be also equipped with two detectors having partial 
azimuthal coverage:
the HMPID (High Momentum Particle Identification Detector)
and the PHOS (Photon Spectrometer).
Besides, an Elecromagnetic Calorimeter (EMCAL) has recently been added to 
the ALICE central setup, covering the 
pseudorapidity range $|\eta|<0.7$ and 110$^\circ$ in azimuth~\cite{EMCAL}.
To be noted that, due to the late inclusion of the EMCAL in the ALICE
design, its installation will be slightly delayed with respect
to the other central detectors.
In the pseudorapidity region $-4.0 < \eta < -2.5$, instead, the ALICE 
Muon Spectometer will be the detector encharged of studying both heavy 
quark vector mesons and 
the $\phi$ meson via the $\mu^+\mu^-$ decay channel, and the production of 
open heavy f\mbox{}lavours. At large rapidity values, other detectors
will be installed, namely the Zero 
Degree Calorimeter (ZDC), the Photon Multiplicity Detector (PMD), 
the Forward Multiplicity Detector (FMD) and the V0 and T0 detectors. 
Finally, the ALICE triggering on cosmic rays will be performed 
by the ACORDE detector (for more details see~\cite{PPRI}). 

\begin{table}[t]
\caption{LHC expected running conditions for pp and PbPb collisions. 
Due to the limited ALICE rate capability, in the experiment interaction
region a lower pp luminosity value is foreseen 
(quoted within brakets in the first row, third column of the Table).}
\vspace{0.4cm}
\begin{center}
\begin{tabular}{|c|c|c|c|c|}
\hline
Collision System & $\sqrt{s_\text{NN}}$ (TeV) & L$_0$ (cm$^{-2}$s$^{-1}$) & Run Time (s/year) & $\sigma_{\text{geom}}$ (b) \\
\hline
pp & 14.0 & $10^{34}$ ($10^{31}$) & $10^7$ & 0.07 \\
PbPb & 5.5 & $10^{27}$ & $10^6$ & 7.7 \\
\hline
\end{tabular}
\end{center}
\label{runningCond}
\end{table}

\section{ALICE Physics Performance}
The ALICE experiment will study 
a wide number of observables, which will require the use of variuos 
experimental techinques\cite{PPRII}. 
In the following, the ALICE expected performance in terms of 
measurement of heavy flavour (Sect.~\ref{sec:open}), quarkonia 
(Sect.~\ref{sec:quarkonia}) and jet production (Sect.~\ref{sec:jet})
will be briefly overviewed.  

\subsection{Heavy Flavour Production in the Central Detectors}
\label{sec:open}
Due to their high mass, heavy flavour quarks (charm and beauty) are 
produced at the very early stage of the collision. For this reason, 
the measurement of open charm and open beauty production is one of the main
observables which can trace the initial phase of the collision and
provide information on the possible formation of the Quark Gluon
Plasma. Experimentally,  
the effects of medium modifications on the final states have
shown up in the parton energy loss observed in AA collisions at RHIC, where a 
departure from binary scaling not only  for light charged hadrons,
but also for heavy flavour production (for 
``non-photonic electrons'', in fact, 
which are likely to come from heavy flavour decays) has 
been observed~\cite{HeavyFlavour1}.

At the LHC, detailed studies on heavy flavour production 
will be even more feasible compared to RHIC,
thanks to the high expected yields for open charm and open 
beauty, which are expected to be of the order of $\sim 100$ c$\bar{\text{c}}$ 
and $\sim 5$ $\text{b}\bar{\text{b}}$ in a 
central PbPb collision\footnote{The quoted values for the open charm 
and open beauty production rate have been computed taking into account also 
the shadowing effect.}. To be noted that these analyses 
will be important also because they will be 
used as a reference for quarkonia production studies
(see Sec.~\ref{sec:quarkonia}). 

\begin{figure}
\begin{tabular}{c|c}
\hskip -0.5cm
\rule[0.4cm]{5.8cm}{0.2mm} & \rule[0.4cm]{10cm}{0.2mm}\\
\hskip -0.5cm
\psfig{figure=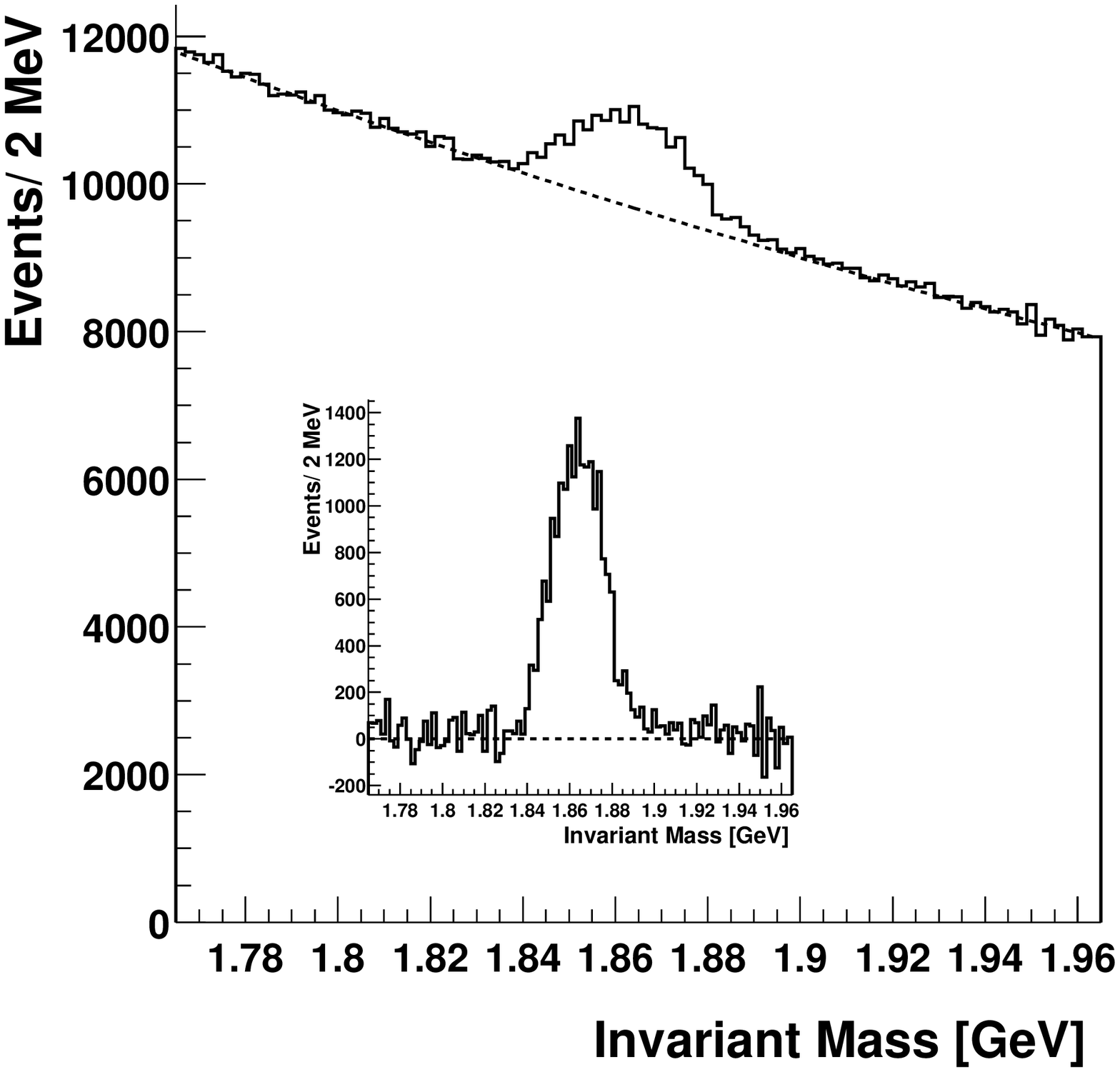,height=5.5cm} & \psfig{figure=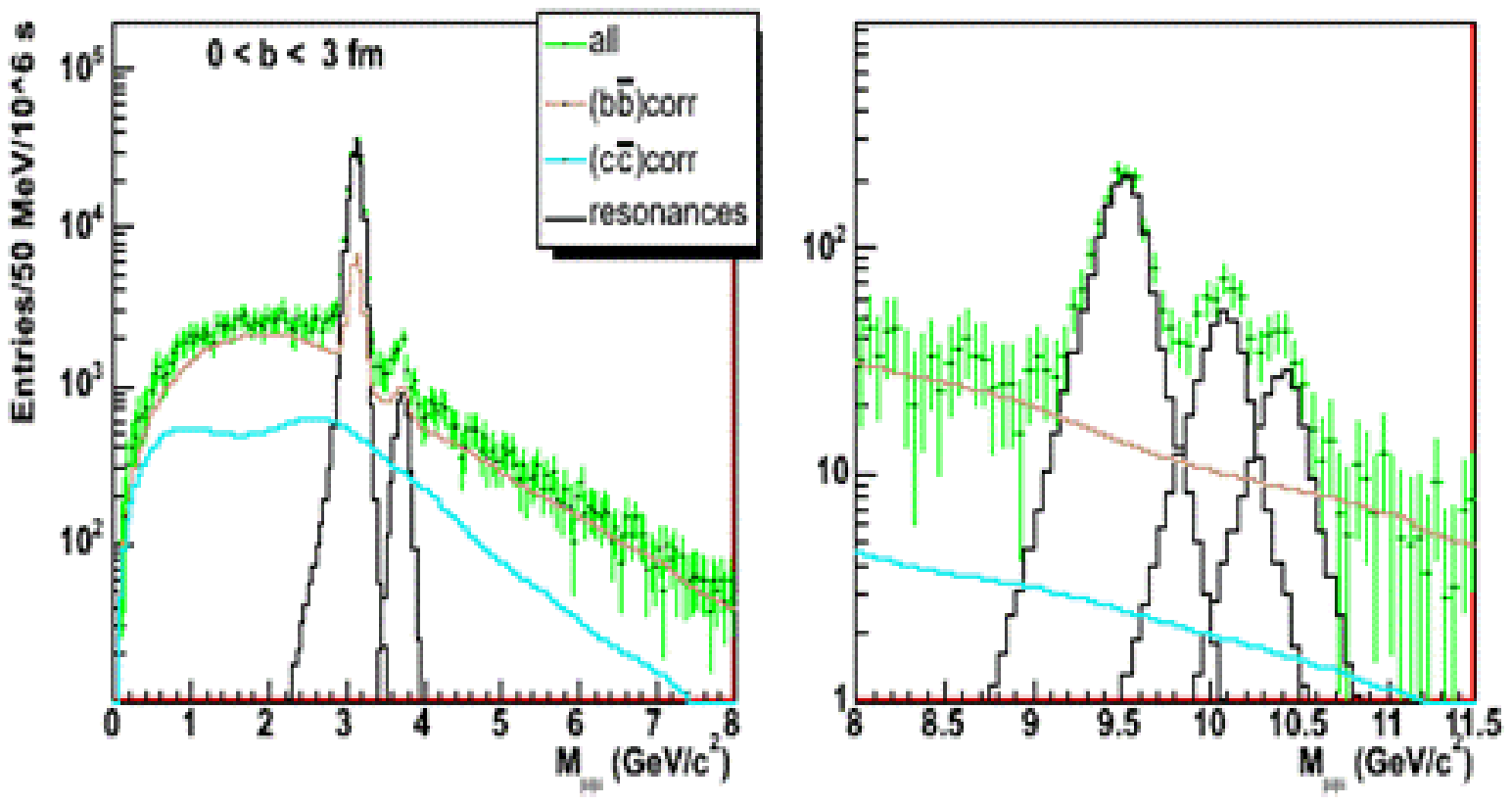,height=5.5cm} \\
\hskip -0.5cm
\rule[-0.2cm]{5.8cm}{0.2mm} & \rule[-0.2cm]{10cm}{0.2mm}\\
\end{tabular}
\caption{Left panel: $\text{K}\pi$ invariant mass distribution for $10^7$ 
PbPb 5\% most central ($b<3.5$ fm) events. The same distribution after 
background subtraction is shown in the inset. Right panel: expected ALICE 
performance for the dimuon spectrum measurement for charmonium (left) and 
bottomonium (right) for a $10^6$s PbPb run (5\% most central events). 
The uncorrelated combinatorial background has been subtracted.}
\label{Kpi}
\end{figure}

The open charm and open beauty analysis will 
rely on the capabilities of the ALICE vertex detector, the ITS, which 
will be characterized by a resolution
on the measurement of the track impact parameter better than 100 $\mu$m
for $\text{p}_\text{T} > 0.6$ GeV/c, allowing to fully reconstruct heavy 
flavour decays. The tracking and the momentum measurement will be
provided by the TPC detector, while particle identification will 
be performed by the TOF detector in the case of charged hadrons, and by the
TRD and the TPC in the case of electrons. 
As an example of the ALICE performance in measuring open charm, 
the left panel of figure~\ref{Kpi} shows the $\text{K}\pi$ invariant mass 
distribution for the study of the $\text{D}\rightarrow \text{K}^-\pi^+$ 
channel, as obtained for $10^7$ PbPb central events\footnote{The 
centrality class of an event has been determined according to the 
impact parameter $b$. In particular, the PbPb 5\% most central events 
correspond to a selection on the event impact parameter $b<3.5$ fm.}.

Being able to measure separately charm and 
beauty hadrons will offer ALICE the opportunity to study the
dependence of the energy loss on the mass of the heavy parton. 
Besides, the heavy-to-light ratio 
$R_{\text{D/h}} (\text{p}_\text{T})$, defined as the ratio of the nuclear 
modification factor of D mesons to that of charged light-flavoured hadrons,
will allow to shed light on the colour-charge dependence of QCD energy loss.
For more details on open charm and open beauty analyses with the ALICE 
detector, see~\cite{PPRII}.

\subsection{Quarkonia}
\label{sec:quarkonia}
The study of quarkonia production in heavy ion collision is one of the 
main observables which can be used 
to investigate the properties of the medium created
in the collision. As a matter of fact, quarkonia are expected 
to be sensitive to the collision dynamics (both at short and long
timescales) and to plasma formation. Moreover, since different quarkonia 
disassociate at different temperatures, they can serve as a 
thermodynamic probe of the medium~\cite{Satz}.

From an experimental point of view, 
the expected anomalous suppression of the $J/\Psi$ related to Debye
colour screening~\cite{JPsi} observed 
at the CERN SPS~\cite{JPsiSPS} was thought to be even stronger at RHIC, 
even in case some dissociation mechanisms due to comoving hadrons would have
occurred. Contrarily, the level of $J/\Psi$ suppression was 
found to be the same as that at the SPS~\cite{JPsiPHENIX}, which 
would imply the presence of some recombination effect~\cite{JPsiReco}
competing with the suppression one. As a consequence, at 
the LHC, because of the even higher   
c$\bar{\text{c}}$ rate, the regeneration mechanism
is expected to dominate, so that
an enhancement in the $J/\Psi$ yield may even take place~\cite{JPsiEnhance}.

The measurement of quarkonia in ALICE will be performed both at midrapidity
in the dielectron channel (making use of the ITS and TPC for tracking, and of
the TRD for electron identification), and in 
the forward rapidity regions, where the ALICE Muon Arm will study the dimuon 
channel. This implies that the ALICE detector will be able to measure 
quarkonia production in two complementary Bjorken-x regions, 
allowing to investigate the PDFs 
in the nuclei. 
Moreover, the feed-down from B decays for $J/\Psi$ production will be kept
under control thanks to the open beauty measurements presented in 
Sec.~\ref{sec:open}. 

The right panel of figure~\ref{Kpi} shows the ALICE 
expected performance in a $10^6$s PbPb
run (5\% most central events) for the measurement of the
dimuon spectrum for charmonium ($J/\Psi$, left) and bottomonium 
($\Upsilon$, right). It has been shown~\cite{PPRII} that for the $J/\Psi$ 
the statistics
collected during one PbPb data taking period will be enough to measure also 
the $\text{p}_\text{T}$ dependence of the charmonium spectrum, while for the 
$\Upsilon$ case, because of the smaller statistics, 
either different centrality classes of events will have 
to be merged, or two or three data taking periods will be necessary. Finally,
it is worth to mention that for quarkonia studies 
the sensitivity of the results to different suppression scenarios has been
investigated. For more details, see~\cite{PPRII}.  

\subsection{Jet Physics}
\label{sec:jet}
Proton-proton collisions at the LHC will be characterized by a very high 
jet production rate. On one hand, jets with 
energy larger than 20 GeV are expected to occur with a frequency of 
$\sim 1$ per PbPb central event; on the other, $\sim 10^5$ highly energetic 
jets with $\text{E}_{\text{T}} \sim 200$ GeV are expected to be produced 
in $10^6$s of PbPb data taking. Thanks to their high transverse energy,
it will be possible to single out and reconstruct these high energy jets 
on an event-by-event basis also in a very entangled environment 
as that produced in a PbPb event. Moreover, the addition of the 
Electromagnetic Calorimeter detector (EMCAL) in ALICE will  
improve the performance of the experiment 
in terms of jet physics, both at the level
of energy measurement, and at the level of the high energy jet triggering 
capabilities of ALICE. 

\begin{figure}
\hskip 3.5cm
\psfig{figure=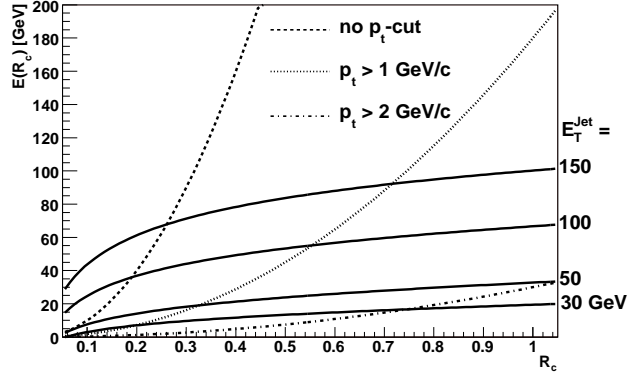,height=5.5cm}
\caption{Charged jet energy within a cone of size R$_\text{c}$ 
(full lines) compared to the energy of the underlying event (dashed lines), 
for different transverse momentum cuts. The background 
energy has been calculated using HIJING quenched  
with $b < 5$ fm.}
\label{jet}
\end{figure}

One of the main characteristics of ALICE jet reconstruction is that
in order to keep under control as much as possible the background level,
it will make use of a limited jet cone size,   
$\text{R}_\text{c} = \sqrt{(\Delta \eta)^2 + (\Delta \phi)^2} 
\sim 0.3 \div 0.4$. 
Moreover, appropriate $\text{p}_\text{T}$ cuts will have to be applied, as 
shown in Figure~\ref{jet}. Here, 
the energy of the jet from charged particles 
as a function of the cone size R$_\text{c}$ 
is drawn (for different energies). The background energy is also plotted for 
different transverse momentum thresholds, from 0 to 2 GeV/c.
As one can see, while the background energy, which is proportional to 
R$_\text{c}^2$, can be reduced either reducing R$_\text{c}$, or by 
applying a p$_\text{T}$ cut (or both), the signal energy is affected 
by these choices only to a minor extend. 

\section{ALICE ``First Day'' Physics}
\label{sec:firstday}
After the collider closure in late 2007, the first
commissioning run at $\sqrt{\text{s}} = 900$ GeV is foreseen to occur 
at the beginning of 2008, when the ALICE setup will be characterized by the 
installation of the complete ITS, TPC, HMPID, Muon Arm, PMD, trigger 
detectors, and of partial configuration of the TOF, TRD and PHOS. 
Already with the first $\sim 10^4$ events collected 
during this run, some ``First Day'' Physics measurements will be feasible, 
namely the study of $\text{dN}/\text{d}\eta$ distributions, $\text{p}_\text{T}$
spectra, multiplicity distributions, and baryon transport analysis. 

To be noted that pp runs at both the commissioning energy 
$\sqrt{\text{s}} = 900$ GeV and at the full-energy 
$\sqrt{\text{s}} = 14$ TeV (in late 2008) will serve not only as a 
baseline for the future ALICE heavy-ion program, but will also be
important by themselves. As a matter of fact, 
the excellent expected ALICE pp performance
in terms of tracking and particle identification,
especially in the low $\text{p}_\text{T}$ range, will make it complementary to
the other LHC proton-proton experiments.

\section{Conclusions}
The ALICE experiment is characterized by a remarkable versatility in terms
of the observables it will look at, and of the experimental techniques it will 
make use of. As the first commissioning run scheduled for the beginning
of 2008 is approaching, the ALICE installation and commissioning is underway, 
and the experiment is evaluating its physics reach with respect to
a wide variety of topics. In this report, to give some examples of the
ALICE expected performance, three selected observables have been discussed, 
namely heavy quarks, quarkonia and jet physics.
  
\section*{Acknowledgments}
The author aknowledges F.Antinori, S. Arcelli and H.A.Gustafsson for their
precious help. Moreover, the author thanks the European Union "Marie Curie" 
Programme for their support for participating at XLII Rencontres de Moriond 
the Conference.


\section*{References}
\begin{thebibliography}{99}

\bibitem{EMCAL} ALICE Addendum to the Technical Proposal: Electromagnetic
Calorimeter, CERN-LHCC-2006-014. 

\bibitem{PPRI} ALICE Collaboration, ALICE: Physics Performance Report Volume
I (2003) CERN/LHCC 2003-049, Eds. F.~Carminati et al. (ALICE Collaboration),
J. Phys. G: Nucl. Part. Phys. 30 (2004) 1517. 

\bibitem{PPRII} ALICE Collaboration, ALICE: Physics Performance Report Volume
II (2005) CERN/LHCC 2005-030, Eds. B.~Alessandro et al. (ALICE Collaboration),
J. Phys. G: Nucl. Part. Phys. 32 (2006) 1295.

\bibitem{HeavyFlavour1} S.S. Adler et al. (PHENIX Collaboration), Phys. Rev.
Lett. 96 (2006) 032301;

B.I. Abelev et al. (STAR Collaboration), nucl-ex/0607012.

\bibitem{Satz} H. Satz, J. Phys. G32 (2006) R25.

\bibitem{JPsi} T. Matsui and H. Satz, Phys. Lett. B178 (1986) 416.

\bibitem{JPsiSPS} M.C. Abreu et al. (NA50 Collaboration) Phys. Lett. B410 
(1997) 337.

\bibitem{JPsiPHENIX} A. Bickley (PHENIX Collaboration) nucl-ex/0701035.

\bibitem{JPsiReco} P. Braun-Munzinger and J. Stachel, Nucl. Phys. A690 (2001)
119;

R.L. Thews et al., Phys. Rev. C63 (2001) 054905;

L. Grandchamp and R. Rapp, Nucl. Phys. A709 (2002) 415.

\bibitem{JPsiEnhance} A. Andronic et al., nucl-th/0611023.

\end{thebibliography}
\end{document}